# Coherent Harmonic Focusing and the Light Extreme

S. Gordienko[1,2], A. Pukhov[1], O. Shorokhov[1], and T. Baeva[1]

[1]*Institut für Theoretische Physik I, Heinrich-Heine-Universität Düsseldorf, D-40225, Germany*
[2]*L. D. Landau Institute for Theoretical Physics, Moscow, Russia*
(Dated: June 7, 2004)

We demonstrate analytically and numerically that focusing of high harmonics produced by the reflection of a few femtosecond laser pulse from a concave plasma surface opens a new way towards unprecedentedly high intensities. The key features allowing for boosting of the focal intensity is the harmonic coherency and the small exponent of the power-law decay of the harmonic spectrum. Using the similarity theory and direct particle-in-cell simulations we find that the intensity at the Coherent Harmonic Focus (CHF) scales as $I_{\rm CHF} \propto a_0^3 I_0$, where $a_0$ and $I_0 \propto a_0^2$ are the dimensionless relativistic amplitude and the intensity of the incident laser pulse. The scaling suggests that due to the CHF, the Schwinger intensity limit can be achieved using lasers with $I_0 \approx 10^{22}$ W/cm$^2$. The pulse duration at the focus scales as $\tau_{\rm CHF} \propto 1/a_0^2$ and reaches the subattosecond range.

The advent of the Chirped Pulse Amplification (CPA) technique [1] was the technological breakthrough leading to a dramatic increase in the achievable laser powers and intensities [2]. The higher laser intensities mean also the higher laser electric fields. As a result, the continuous technological progress has allowed for laboratory studies of the ever new physics, from the strong field laser-atom interactions [3], up to the ultra-relativistic laser-plasmas [4] and the high-energy particle acceleration [5]. This physics is of interest, because it studies non-linear properties of matter in the ultra-strong laser fields.

One may ask a question, whether the intensity can be reached that allows studying non-linear properties of one of the most intriguing media: of vacuum itself. It is known, that vacuum is in fact not empty. It is filled with virtual particles. The lightest charged particles participating in the electromagnetic interactions are electrons and positrons. They continuously appear and disappear in vacuum on distances of the order of the Compton wavelength $\lambda_C = \hbar/mc$. Quantum electrodynamics suggests that there exists a very fundamental critical value of the electric field: $E_{\rm QED} = m^2 c^3/e\hbar = 1.3 \times 10^{16}$ V/cm, here $m$ and $-e$ are the electron mass and charge respectively. An electric field with the strength $E_{\rm QED}$ accomplishes the work $eE_{\rm QED}\lambda_C = mc^2$ on the Compton distance and thus is able to bring these pairs from the virtual realm into the real world. The vacuum gets polarized and its response becomes highly non-linear. The critical field $E_{\rm QED}$ is known as the Schwinger limit [6]. The corresponding laser intensity $I_{\rm QED}$ can be easily calculated: $I_{\rm QED} = cE_{\rm QED}^2/4\pi = 4.6 \times 10^{29}$ W/cm$^2$. To realize, how huge this value is, let us suppose one wishes to reach the $I_{\rm QED}$ intensity by focusing an ultra-short laser pulse down to the $\lambda^3$ volume [7], where $\lambda$ is the laser wavelength. Then, the laser energy must be $W = 4\pi I_{\rm QED} \lambda^3/3c = 6.4 \cdot 10^7$ J$(\lambda/\mu{\rm m})^3$. For a laser with $\lambda \approx 1\mu$m this would mean a 64 MJ pulse energy, which is hardly feasible in the foreseen future.

In this work we present the concept of a Coherent Harmonic Focusing (CHF) and demonstrate by means of theoretical analysis and direct particle-in-cell (PIC) simulations that the Schwinger limit can be achieved at reasonable laser pulse energies using the existing or just emerging laser technology.

The basic idea of harmonic focusing is to take an initial laser pulse with the wavelength $\lambda_0$, send it through a nonlinear medium, generate $n$ high harmonics with the wavelengths $\lambda_n = \lambda_0/n$ and then focus them down to a spot size $\sim \lambda_n$. However, one has to distinguish between coherent and incoherent harmonics. If the harmonics are incoherent, then the harmonics intensities are to be added. Since the dimension of the focal spot scales as $1/n$, the field at the incoherent focus is boosted only if the harmonic spectrum decays slower than $1/n^2$.

The case of CHF is very much different. One generates high harmonics coherently and focuses them in such a way that the fields of *all* harmonics interfere constructively within the (very small!) focal volume. We see later that in order to boost the intensity by means of the CHF mechanism, the harmonic spectrum must decay slower than $1/n^4$. Such spectra do exist. Recently, it has been shown that the high harmonics spectrum produced in the laser interaction with a sharp plasma boundary is a universal one and it decays as $1/\omega^{5/2}$ [8]. It is also important that the laser-plasma surface harmonics are coherent and appear in the form of (sub-)attosecond pulses [7, 8]. Experimentally, the plasma surface harmonics are produced by irradiating the surface of a solid material by a relativistically intense laser pulse [9]. Being exposed to the laser, the surface becomes a plasma with the solid state density. Shaping the target surface appropriately, one can focus the harmonics.

To explain the CHF mechanism let us consider a laser wave with the vector potential

$$\mathbf{A}(t,x) = \mathbf{A}_0 \exp\left(-\frac{(x/c-t)^2}{\tau^2} + i\omega_0(x/c-t)\right) + {\rm c.c.}$$

This wave is reflected by a sharp surface of the plasma electron fluid. The ponderomotive force of the laser pulse causes the reflecting surface to oscillate back and forth with relativistic velocities. The reflected radiation contains high harmonics [8].

We are interested in the reflection from an overdense plasma $N_e \gg N_c$ with $\Gamma = a_0 (N_c/N_e) \ll 1$, where $a_0 = eA_0/mc^2$ is the dimensionless vector potential, $N_e$ is the plasma electron density, and $N_c = \omega_0^2 m/4\pi e^2$ is the critical density. The reflected radiation can be expressed as the Fourier integral

$$\mathbf{E}_r(t, \mathbf{r}) = \int_0^{+\infty} \mathbf{E}_\omega \exp(i\omega t + i\omega x/c)\, d\omega + \text{c.c.}, \quad (1)$$

where $\mathbf{r} = (x, y, z)$. According to the work [8] the spectrum of the reflected radiation is

$$|\mathbf{E}_\omega|^2 = \eta c^{-2} A_0^2 (\omega_0/\omega)^p, \quad \arg \mathbf{E}_\omega \approx C\omega\tau + \varphi + \pi/2 \quad (2)$$

for $1 \ll \omega/\omega_0 \leq n_c$, where $n_c = 4\gamma_{max}^2$, $\gamma_{max}$ is the largest relativistic factor of the reflecting surface, $\eta$ is the conversion efficiency, $\varphi$ is the initial harmonics phase, $C$ is a constant. The exponent $p = 5/2$ or $p = 3$ depending on the interaction regime [8].

Eq. (1) is written for a plane wave reflected from a plane surface. To treat reflection from a curved surface, we re-write (1) using the Huygens principle [10]:

$$\mathbf{E}_r(t, \mathbf{r}) = \int_0^{+\infty} \mathbf{E}(\omega, \mathbf{r}) \exp(i\omega t)\, d\omega + \text{c.c.}, \quad (3)$$

where

$$\mathbf{E}(\omega, \mathbf{r}) = \frac{\omega}{2\pi i c} \int \frac{\exp(-i\omega R/c)}{R} \mathbf{E}(\omega, \mathbf{r}')\, dS \quad (4)$$

and the integral is taken over the wave front $S$. Here, $\mathbf{E}(\omega, \mathbf{r}')$ is the Fourier component of the electric field at the point $\mathbf{r}'$ of $S$, $R = |\mathbf{r} - \mathbf{r}'|$.

As an example we consider the simplest case when a spherical wave is reflected from a segment of a co-focal spherical surface with the radius $R_0$. The segment occupies the solid angle $\Omega \ll 4\pi$. If $R_0 \gg \lambda$, then the spectrum (2) is valid at every reflection point, and the focal field is

$$\mathbf{E} = R_0 \Omega \left[ \int \omega \mathbf{E}_\omega \exp\left(i\omega\left(t - \frac{R_0}{c}\right)\right) \frac{d\omega}{2\pi c i} \right] + \text{c.c.} \quad (5)$$

Notice that the multiplier $\omega$ appears due to focusing provided by the spherical geometry.

Substituting the power-law spectrum (2) into the integral in (5), we find that the field reaches its maximum at the focus at $t \approx t_f = R_0/c - C\tau$:

$$\frac{|\mathbf{E}_f|^2}{|\mathbf{E}_0|^2} = \eta \left(\frac{4R_0\Omega}{\lambda}\right)^2 \left(\frac{n_c^q - 1}{4 - p}\right)^2 \cos^2 \varphi. \quad (6)$$

Here we have defined $q = 2 - p/2$ and $\mathbf{E}_0 = \omega_0 \mathbf{A}_0/c$. Note that for $p > 4$ one has $q < 0$ and $n_c^q \ll 1$. In this case, $|\mathbf{E}_f|^2$ is defined by low order harmonics. However, for $p < 4$ we have $q > 0$ and $n_c^q \gg 1$ that leads to

$$\frac{|\mathbf{E}_f|^2}{|\mathbf{E}_0|^2} = \eta \left(\frac{4R_0\Omega}{\lambda}\right)^2 \frac{n_c^{2q}}{(4 - p)^2} \cos^2 \varphi, \quad (7)$$

In this case $|\mathbf{E}_f|^2$ is defined by the coherent focusing of high order harmonics and the CHF intensity boosting factor is $n_c^{2q}$. The oscillating integral in (5) gives the pulse duration at the focus $\tau_f = 2\pi/(\omega_0 n_c)$.

To highlight the importance of harmonics coherency let us assume a general power-law spectrum of the electric field $\mathbf{E}_\omega \propto \exp(-i\Psi(\omega))/\omega^{p/2}$. For such a spectrum one finds that the intensity at the focus is

$$|\mathbf{E}|^2 \propto \left| \Re \int \frac{\exp[-i\Psi(\omega) + i\omega(t - R_0/c)]}{\omega^{p/2-1}} d\omega \right|^2. \quad (8)$$

If the harmonics are incoherent, then the function $\Psi(\omega)$ is a fast oscillating one and only one harmonic with the frequency satisfying $d\Psi(\omega)/d\omega = t - R_0/c$ significantly contributes to the integral in (8) at the time $t$. Thus there is no increase in the intensity at the focus due to the incoherent harmonic focusing for $p > 2$.

Now we are going to derive an analytical scaling for the CHF-intensity in the focal spot as a function of the incident laser amplitude and the plasma density. To tackle this problem we have to solve the Vlasov equation on the electron distribution function $f(t, \mathbf{r}, \mathbf{p})$

$$[\partial_t + \mathbf{v}\partial_\mathbf{r} - e(\mathbf{E} + \mathbf{v} \times \mathbf{H}/c)\partial_\mathbf{p}] f(t, \mathbf{p}, \mathbf{r}) = 0, \quad (9)$$

together with the Maxwell equations on the electric $\mathbf{E}$ and magnetic $\mathbf{H}$ fields. A dimension analysis yields

$$f = \frac{N_e}{(mc)^3} F\left(\omega_0 t, \frac{\mathbf{p}}{mc}, \frac{\omega_0 \mathbf{r}}{c}, \frac{N_c}{N_e}, a_0, \omega_0 \tau\right), \quad (10)$$

where $F$ is an unknown universal function. Eq. (10) is of little use as long as it depends on three dimensionless parameters. However, we are interested only in the ultrarelativistic limit. Thus, we can set $\mathbf{v} = c\mathbf{n}$, $\mathbf{n} = \mathbf{p}/|\mathbf{p}|$ and re-write the Vlasov equation as

$$[\partial_t + c\mathbf{n}\partial_\mathbf{r} - e(\mathbf{E} + \mathbf{n} \times \mathbf{H})\partial_\mathbf{p}] f = 0. \quad (11)$$

Further, we introduce the dimensionless variables $\hat{t} = \omega_0 t$, $\hat{\mathbf{p}} = \mathbf{p}/mca_0$, $\hat{\mathbf{r}} = \omega_0 \mathbf{r}/c$, $(\hat{\mathbf{E}}, \hat{\mathbf{H}}) = c(\mathbf{E}, \mathbf{H})/\omega_0 A_0$ and re-write Eq. (11) together with the Maxwell equations in the dimensionless form

$$\left[\partial_{\hat{t}} + \mathbf{n}\partial_{\hat{\mathbf{r}}} - e\left(\hat{\mathbf{E}} + \mathbf{n} \times \hat{\mathbf{H}}\right)\partial_{\hat{\mathbf{p}}}\right] \hat{f} = 0;$$
$$\nabla_{\hat{\mathbf{r}}} \cdot \hat{\mathbf{E}} = (1 - \hat{\rho})/\Gamma, \quad \nabla_{\hat{\mathbf{r}}} \cdot \hat{\mathbf{H}} = 0, \quad (12)$$
$$\nabla_{\hat{\mathbf{r}}} \times \hat{\mathbf{H}} = \hat{\mathbf{j}}/\Gamma + \partial_{\hat{t}}\hat{\mathbf{E}}, \quad \nabla_{\hat{\mathbf{r}}} \times \hat{\mathbf{E}} = -\partial_{\hat{t}}\hat{\mathbf{H}},$$

where $\hat{\rho} = \int \hat{f} d\hat{\mathbf{p}}$, $\hat{\mathbf{j}} = \int \mathbf{n} \hat{f} d\hat{\mathbf{p}}$. Eqs. (12) contain the only one dimensionless parameter $\Gamma = a_0 N_c/N_e$ and the unknown universal function $\hat{f}(\hat{t}, \hat{\mathbf{p}}, \hat{\mathbf{r}}) = (m^3 c^3 a_0^3/N_e) f(t, \mathbf{p}, \mathbf{r})$. Now we write for the distribution function

$$f = \frac{N_e}{(mca_0)^3} \hat{f}\left(\omega_0 t, \frac{\mathbf{p}}{mca_0}, \frac{\omega_0 \mathbf{r}}{c}, \Gamma, \omega_0 \tau\right). \quad (13)$$

It follows immediately from (13) that the relativistic $\gamma$-factor of the reflecting surface scales as $\gamma(t) = a_0 \hat{\gamma}(\omega_0 t, \omega_0 \tau, \Gamma)$, where $\hat{\gamma}$ is a universal function. As a result, one finds

$$\gamma_{max} = G(\Gamma, \omega_0 \tau) a_0, \quad n_c = 4 a_0^2 G^2(\Gamma, \omega_0 \tau) \quad (14)$$

as well as $\eta = \eta(\Gamma, \omega_0 \tau)$, $\varphi = \varphi(\Gamma, \omega_0 \tau)$, where all the functions $\eta$, $G$ and $\varphi$ are universal.

Having at our disposal the similarity theory we can find the focal intensity analytically. We choose the parameters $\omega_0 \tau$ and $\Gamma$ so that the spectral slope $p = 5/2$ and $q = 3/4$. Then, the CHF amplification factor is $n_c^{3/2}$. Let us specify our results for this particular case. From Eq. (6) we obtain a scaling for the focal intensity $I_{\mathrm{CHF}}$ produced by the CHF and for the pulse duration at the focus $\tau_{\mathrm{CHF}}$. If one fixes the dimensionless parameter $\Gamma$ and changes the laser amplitude $a_0$ together with the plasma density $N_e$ in such a way that $\Gamma = a_0 N_c/N_e = const$, then

$$I_{\mathrm{CHF}} = \mu_1 (R_0 \Omega/\lambda)^2 a_0^3 I_0, \quad \tau_{\mathrm{CHF}} = 2\pi \mu_2/(a_0^2 \omega_0), \quad (15)$$

where $I_0$ and $a_0$ are the incident pulse intensity and its dimensionless amplitude at the reflecting surface; $\mu_1 = \mu_1(\omega_0 \tau, \Gamma)$ and $\mu_2 = \mu_2(\omega_0 \tau, \Gamma)$ are universal functions with their values on the order of unity.

It follows from (15) that the Schwinger limit at the coherent harmonic focus can be reached for the incident laser pulse intensity

$$I_{crit} = \left(\frac{\lambda}{R_0 \Omega \sqrt{\mu_1}}\right)^{4/5} \left(\frac{\hbar \omega_0}{mc^2}\right)^{6/5} I_{\mathrm{QED}}. \quad (16)$$

Assuming the geometrical factor $R_0 \Omega \sqrt{\mu_1}/\lambda \approx 1$, we get $I_{crit} \approx 8.5 \cdot 10^{22} \, (\mu m/\lambda)^{6/5}$ W/cm$^2$.

To demonstrate the CHF principle, we have done direct particle-in-cell (PIC) simulations using the code Virtual Laser-Plasma Laboratory (VLPL) [11]. In the three-dimensional simulations we take a linearly polarized spherical laser wave reflecting from a co-focal spherical mirror. The laser pulse has a Gaussian temporal profile: $a(t, R) = a_0 (R_0/R) \exp(-t^2/T^2) \cos(\omega_0(t - R/c))$ with the amplitude $a_0 = 3$ when it arrives at the mirror surface located at $R_0 = 4\lambda$. The plasma has the density $N = 5 N_c$. The pulse duration was $T = 2\pi/\omega_0$. To compare the CHF and a simple geometric focusing of the laser fundamental wave we have done another simulation, where the spherical laser wave was converging down to the theoretically smallest possible spot size $\lambda_0/2$.

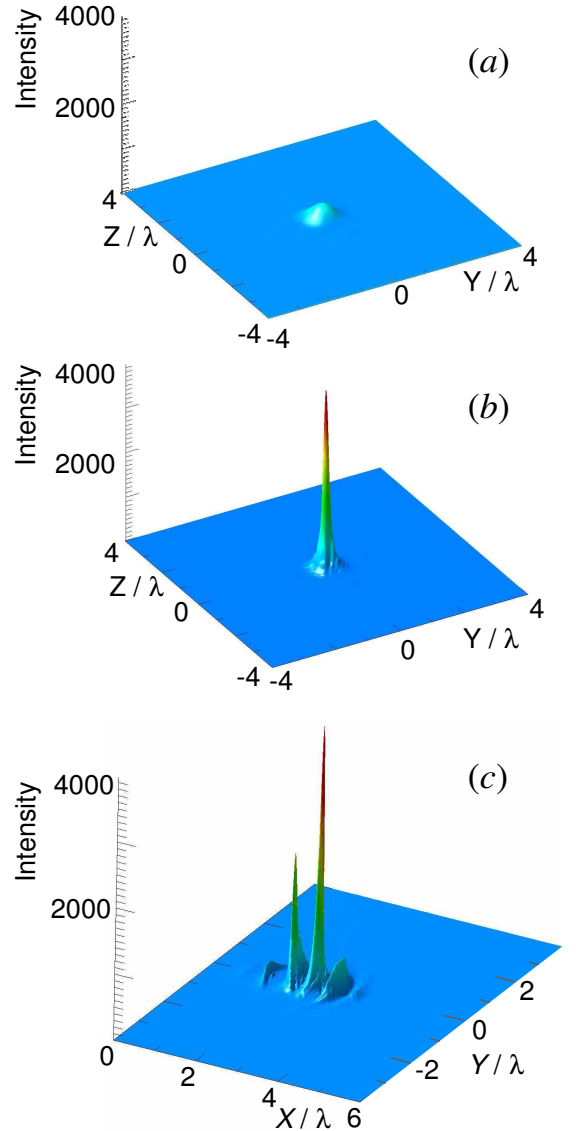

FIG. 1: (color). 3D PIC simulation results, distribution of the dimensionless intensity $I = (E^2 + B^2)(e^2/2mc\omega_0)^2$. (a) Intensity distribution in the focal plane $(YZ)$ due to simple focusing of the laser fundamental; (b) intensity amplification in the focal plane $(YZ)$ by the CHF effect; (c) on-axis CHF intensity cut in the polarization plane $(XY)$: the periodic structure is defined by the laser fundamental. The characteristic very sharp intensity spike in the focus is due to the CHF boosting.

The simulation results are presented in Fig. 1. The frame Fig. 1a shows the intensity distribution in the focal plane of the converging fundamental laser wave (no harmonics). As a contrast, Fig. 1b shows the focal plane of the CHF produced by the laser wave bounced off a concave plasma surface. The intensity in the center is boosted by more than an order of magnitude as compared with the simple linear focusing. Fig. 1c shows the on-axis cut of the CHF reflected intensity in the polarization plane $(XY)$ at the focusing time. Here one sees a periodic structure defined by the half-wavelength of the laser fundamental and the very sharp intensity spike at the CHF focus. The same spike is perfectly seen also in Fig. 1b. The spike tip has a width of the single grid cell $h_y = h_z = 0.0125\lambda_0$ in our 3D PIC simulation. This means that the numerical grid was insufficient to resolve the CHF reliably. Yet, the 3D PIC simulation contained already $3 \times 10^8$ grid cells and $10^9$ numerical particles. This is the very limit of the available computational power. The further grid refining by factor 2 in each dimension of the 3D geometry would require 16 times more computational time and 8 times more memory, which is not feasible presently.

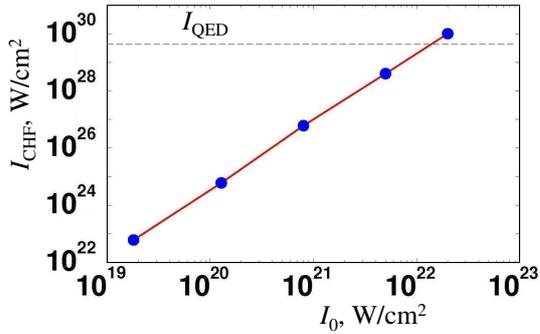

FIG. 2: Numerically obtained scaling for the CHF focal intensity versus the incident laser amplitude. The fundamental laser wavelength is assumed $\lambda_0 = 750$ nm, the dimensionless parameter $\Gamma = aN_c/N_e = 0.6$. The broken line marks the vacuum breakdown intensity $I_{\text{QED}}$. The numerical scaling agrees with the analytical result (15).

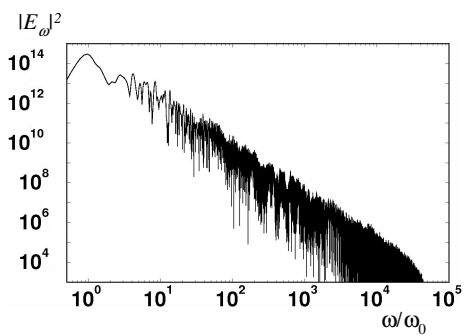

FIG. 3: Harmonics spectrum $|E_\omega|^2$ of the reflected radiation for the incident laser intensity $2.4 \times 10^{22}$ W/cm$^2$ and the relativistic amplitude $a_0 = 100$. The power-law spectrum $|E_\omega|^2 \propto (\omega_0/\omega)^{5/2}$ reaches up to the frequency $\omega \approx 5 \times 10^4 \omega_0$.

To fit the problem into the available computational resources, we used 1D PIC simulations and assumed that the 1D harmonics are reflected by a spherically focusing mirror. Then, we applied the operator (5) to the harmonics $\mathbf{E}_\omega$ taken from the 1D PIC results. The mirror radius was $R_0 = 4\lambda_0$ and the solid angle $\Omega = 1$. On this way we were able to obtain numerically the scaling for the CHF focal intensity $I_{\text{CHF}}$ over a wide range of the incident laser intensities $I_0$. The results are shown in Fig. 2. We have assumed the fundamental laser wavelength $\lambda_0 = 750$ nm and fixed the dimensionless parameter $\Gamma = aN_c/N_e = 0.6$. The numerical results in Fig. 2 agree well with the analytical scaling (15). The broken line in Fig. 2 marks the vacuum breakdown intensity $I_{\text{QED}}$. The scaling in Fig. 2 shows that the intensity $I_{\text{QED}}$ can be achieved in the CHF focus by using an incident laser pulse with $I_0 \approx 10^{22}$ W/cm$^2$.

The highest incident laser pulse intensity we have simulated in the 1D (the rightmost upper point in Fig. 2) was $2.4 \times 10^{22}$ W/cm$^2$ corresponding to the relativistic amplitude $a_0 = 100$. We present the reflected radiation spectrum $|E_\omega|^2$ in Fig. 3. One sees that the power-law spectrum $|E_\omega|^2 \propto (\omega_0/\omega)^{5/2}$ reaches up to the frequency $\omega \approx 5 \times 10^4 \omega_0$. It is this slow-decaying harmonics spectrum that allows one to achieve the extremely high intensity via the CHF boosting.

In conclusion, we propose a new way to achieve extreme intensities in the coherent harmonics focus. The CHF effect allows to reach the Schwinger limit of vacuum polarization using source laser pulses with reasonable intensities. Simultaneously, the CHF works as a spectral filter and shortens the pulse duration down to the zeptosecond range.

We are very obliged to Prof. Gerard Mourou for fruitful discussions on the high harmonics focusing.

This work has been supported in parts by AvH fund, DFG (Germany), and by RFFI 04-02-16972, NSH-2045.2003.2 (Russia).